\documentclass[pra,twocolumn,showpacs,aps]{revtex4}
\usepackage{graphicx}
\usepackage{latexsym}

\begin{document}

\title{Non-linear quantization for arbitrary distributions and applications to Medical Image Processing.}

\author{C. Tannous}

\affiliation{Laboratoire de Magn\'{e}tisme de Bretagne, UPRES A CNRS 6135, Universit\'{e} de Bretagne Occidentale, BP: 809 Brest CEDEX, 29285 FRANCE}
 
\date{November 14, 2002}


\begin{abstract}
We report the development of a scalar quantization approach that helps build tables of decision and reconstruction levels for any probability density function (pdf). Several example pdf's are used for illustration: Uniform, Gaussian, Laplace, one-sided Rayleigh, and Gamma (One sided and  double-sided symmetrical). The main applications of the methodology are principally aimed at Multiresolution Image compression where generally the Stretched Exponential pdf is encountered. Specialising to this important case, we perform quantization and information entropy calculations from selected medical MRI (Magnetic Resonance Imaging) pictures of the human brain. The image histograms are fitted to a
Stretched exponential model and the corresponding entropies are compared.


\end{abstract}

\maketitle

\section{Introduction}

This research attempts at developing an information preserving image coding suitable for handling medical images which usually do not allow any degree of image degradation in comparison with the source images \cite{saghri,takaya}. JPEG, one of the well accepted compression standards, exhibits an excellent compression gain, but does not insure that the decoded result is the exact replica of the original. A variation of DPCM or pixel difference coding, namely "Guided Scan Pixel Difference Coding  was studied. A major modification to the classical DPCM is the inclusion of scan information, in terms of the direction used to calculate the pixel difference and step size. This additional information does not significantly increase the volume of coded data. It actually opens the extensibility for the DPCM to be able to handle multi-dimensional image data, in a similar manner as JPEG is extended to MPEG to handle movie presentation. Another important asp ect of medical image coding, which is the ability to access the full details of a local image (often referred to as ROI, Region of Interest), is also satisfied. Neither the location of a ROI nor the scanning method is restricted so that a local area of ROI can be selected anywhere in the original image. The flexibility in scanning makes it possible to use the same code for real time movie presentation in the same way MPEG uses JPEG. This research considers two stage redundancy removal, one by taking pixel differences and the other by an entropic source coding similar to the Huffman coding to achieve a redundancy removal of approximately 4-5 bpp (bits per pixel).\\

The sizes of medical images have increased with the advancement in various medical imaging modalities, typically in MRI and X-ray CAT scan. MRI can scan a whole body within a reasonable time of 30 minutes and X-ray CAT has adopted helical scanning for a higher spatial resolution. In the case of 3D scanning, the image size could be as large as 100 million pixels. Images of this size are usually stored on writable laser disks (CD-WORM), but this imposes serious problems when the retrieval of such data is attempted through a communication channel in the wide area network (WAN) environment using 64 kbits/sec bit-rate telephone channels.\\

The problems associated with the compression of medical images are two fold. The first requirement is that compression must be absolutely non-degradable. Regardless of whether a picture has been affected by measurement noise that occurs in the process of physical measurement and in image reconstruction, the source data acquired at an imager must be retained without any kind of losses. The second condition is the integrity of the objects contained in a picture data. Physicians examine very closely the image in a ROI (Region of Interest) or a number of ROI's. A selected ROI needs to be immediately accessible and it must be translated into a serial stream of compressed data for transmission via a communication channel.

\section{Lloyd-Max quantization fundamentals}

The scalar quantization problem entails the specification of a set of decision levels $d_{j}$ and a set of reconstruction levels $r_{j}$  in a way such that some quantization error measure is minimized.
The Lloyd-Max algorithm is based on the minimization of the mean-square error defined as:

\begin{eqnarray}
&& \mathcal {E}= E[(x-\hat{x})^2]=\int_{a}^{b} dx \hspace{1mm} p(x) (x-\hat{x})^2  \nonumber \\
&& \hspace{0.5cm} = \sum_{j=0}^{J-1} {\int_{d_{j}}^{d_{j+1}} {dx \hspace{1mm}p(x)(x-r_{j})^2}}
\end{eqnarray} 

Where the probability density function (pdf) $p(x)$ is defined over the interval $[a,b]$. $E$ is the expectation
value with respect to the pdf $p(x)$ and $\hat{x}$ is the mean value with respect to $p(x)$.
The total number of levels $J= 2^{n_{b}}$ depends on the number $n_{b}$ of available bits to encode them.
Considering $ \mathcal {E}$ a function of all the $r_{j}$   and the $d_{j}$   levels, the minimum of the mean-square error is obtained by taking the derivatives of $\mathcal E$ with respect to the $r_{j}$   and the $d_{j}$. This yields the coupled sets of equations $(j=1,...J)$:

\begin{eqnarray}
 &  \frac{ \partial{\mathcal{E} }}{\partial{d_{j}}}= (d_{j}- r_{j})^{2}p(d_{j})- (d_{j}- r_{j-1})^{2}p(d_{j}) =0  \\
 &  \frac{ \partial{\mathcal{E} }}{\partial{r_{j}}}= 2 \int_{d_{j}}^{d_{j+1}}{(x- r_{j})p(x)dx}
\end{eqnarray}

Simplifying, these equations are rewritten as:

\begin{eqnarray}
 &  r_{j} = 2d_{j}- r_{j-1}  \\
 &   r_{j}= \frac{  \int_{d_{j}}^{d_{j+1}} {dx \hspace{1mm}p(x) \hspace{1mm}x }   } {  \int_{d_{j}}^{d_{j+1}} {dx \hspace{1mm}p(x) }  }
\end{eqnarray}

These sets of equations form the Lloyd-Max algorithm.

Starting from some initial value $r_{0}$  the Lloyd-Max set of equations will allow us to calculate all the decision and reconstruction levels. In fact, it appears that the error $\mathcal{E}$ is a function of $r_{0}$  only when we are dealing with the mean-square error quantization.
The pseudo-code algorithm is therefore:
\begin{eqnarray}
&& \mbox{ Start with } d_{0}=a; \nonumber \\
&& \hspace{2cm}	\mbox{ for a given } r_{0} ;  \nonumber \\
&& \hspace{3cm}		\mbox{ find }   min \{ \mathcal{E} (r_{0} ) \} ; \nonumber \\
&& \hspace{4.5cm}			j=1; \nonumber \\
&& \hspace{3cm}				(*) \mbox{ from [5] get } d_{j}; \nonumber \\
&& \hspace{2cm}			\mbox{ from [4] get } r_{j}; \nonumber \\
&& \hspace{1cm}		            j=j+1; \nonumber \\
&& \hspace{0.5cm}	if \hspace{1mm} ( j \le J ) \mbox{ go to } (*); \nonumber \\
&& end
\end{eqnarray}

The above algorithm has two main ingredients:
\begin{itemize}
\item A minimization procedure to find the optimal value of the starting reconstruction level $r_{0}$. 
\item A root-solving procedure allowing the determination of $d_{j+1}$  from $r_{j}$  and $d_{j}$  with an implicit function [5] of $d_{j+1}$  which is the ratio of two integrals.
\end{itemize}

Several difficulties might arise:
\begin{enumerate}
\item The function $ \mathcal{E} (r_{0} )$ is highly singular with many local minima separarted by large values of the function as illustrated in Figure 1. This means that one has to have a good guess for the starting value $r_{0}$  otherwise one will converge to the local minmium and not the global one.
\item The integrals appearing in [2] should be calculated with a very high accuracy despite the fact we are dealing with very rapidly exponential functions in the integrand (this is the case of the Laplace, Gamma, Rayleigh, Gaussian and Stretched exponential pdf's. The Uniform distribution does not have these problems but many other pdf's contain an exponential term).
\end{enumerate}

\section{Implementation of the Algorithm and applications}

We have implemented an algorithm based on an analytical approach that will deal with an integrand containing arbitrary powers of x and an exponential function of an arbitrary power of x.
The minimization procedure is handled graphically within the software by plotting the function $ \mathcal{E} (r_{0} )$ over a large interval, allowing the approximate location of the starting point that will drive the search towards the global minimum region.\\
Several examples will be treated below and look-up tables are given for the levels for a quantization based on an arbitrary number of bits. Some of these tables are compared to highly accurate published tables and new tables are given for the stretched exponential pdf needed in mutiresolution image processing.
The various pdf that are illustrated below are:

\begin{enumerate}

\item Uniform pdf
\begin{eqnarray} 
p(x) & = & 1/2a	\hspace{2cm} x \in \hspace{1mm} [-a, a]. \\
p(x) & = & 0     \hspace{2.5cm}     x \mbox{ elsewhere}.
\end{eqnarray}
							
Mean $\mu=0$, Standard deviation: $\Delta x=a^{2}/3$.

\item Gaussian pdf:

\begin{equation}
p(x)=\frac{1}{\sqrt{2\pi \sigma^{2}}}exp(- \frac{x^2}{2\sigma^2}), \hspace{0.5cm} x \in \hspace{1mm} ]-\infty,+\infty[
\end{equation}

Mean $\mu=0$, Standard deviation: $\Delta x=\sigma$.

\item Laplace pdf:

\begin{equation}
p(x)=\frac{1}{ \sigma \sqrt{2} }exp(- \frac{\sqrt{2} |x| }{\sigma})  \hspace{1cm} x \in \hspace{1mm} ]-\infty,+\infty[
\end{equation}

Mean $\mu=0$, Standard deviation: $\Delta x=\sigma$.

\item Double-sided Gamma pdf:

\begin{equation}
p(x)=\frac{\sqrt[4]{3}}{\sqrt{ 8 \pi \sigma |x| }}exp(- \frac{\sqrt{3} |x| }{ 2 \sigma}), \hspace{0.2cm} x \in \hspace{1mm} ]-\infty,+\infty[
\end{equation}

Mean $\mu=0$, Standard deviation: $\Delta x=\sigma$.

\item Stretched Exponential pdf:

\begin{eqnarray}
&& p(x) = K e^{- {(\frac{|x|}{\alpha})}^\beta}, \nonumber \\
&&  \hspace{1mm} \mbox{ with } K=\frac{\beta}{ 2\alpha\Gamma(\frac{1}{\beta}) }  \mbox{ and } x \in \hspace{1mm} ]-\infty,+\infty[
\end{eqnarray}

Mean $\mu=0$, Standard deviation: $\Delta x= \alpha \sqrt{ \frac{ \Gamma(\frac{3}{\beta})}{\Gamma(\frac{1}{\beta})}}$. $\Gamma(x)$ is the Euler Gamma function.

\item Single-sided Rayleigh pdf:

\begin{equation}
p(x)=\frac{x}{ \sigma^{2} }exp(- \frac{x^{2}}{2 \sigma^{2}}) \hspace{1.5cm} x \in \hspace{1mm} ]0,+\infty[
\end{equation}

Mean $\mu=\sigma \sqrt{\pi /2}$, Standard deviation: $\Delta x=\sigma \sqrt{2(1-\pi/4)}$.

\item One-sided Gamma pdf:

\begin{eqnarray}
&& p(x)=K_{g} x^{\alpha-1}exp( - \frac{x}{\beta} ), \nonumber \\
&& \hspace{1mm} \mbox{ where }  K_{g}=\frac{1}{\beta^{\alpha}\Gamma(\alpha)}  \mbox{ and } x \in \hspace{1mm} ]0,+ \infty[
\end{eqnarray}

Mean $\mu=\alpha \beta$, Standard deviation: $\beta \sqrt{\alpha}$.

\end{enumerate}

Four bit quantization tables of decision and reconstruction levels for all the above double-sided pdf in the case the mean $\mu=0$ and the standard deviation $\Delta x=1$  are given in Tables 1-7 and are compared to published values in Table 8 and 9.\\

For the one-sided Rayleigh and Gamma pdf's we consider the case where the standard deviation $\Delta x=1$ (For the one-sided Gamma pdf we take $\alpha=1.2$). The  $\alpha$ parameter of the stretched exponential pdf with mean  $\mu=0$ and standard deviation  $\Delta x=1$ can be set arbitrarily. We set it to 1.2 as in the one-sided Gamma pdf. 
Figures 2 and 3 illustrate the reconstruction levels versus the decision levels in the case the number of bits is 6. The quantized levels are shown for the Uniform, Gaussian, Laplace, Symmetric Gamma, Stretched exponential and the one-sided Rayleigh and Gamma pdf's. All symmetric pdf's have ($\mu=0$ and $\Delta x=1$) whereas the one-sided pdf's have $\Delta x=1$. 

\section{Multiresolution and Medical Image Processing}

From a histogram of an image pixel intensities, we evaluate the absolute mean $|\mu|$ and standard deviation $\sigma$ and extract from them the parameters $\alpha$ and $\beta$ of the stretched exponential probability density function given by Eq.12.\\

Following [1], we perform firstly the extraction of the parameter $\beta$ from the equation:

\begin{equation}
F(\beta) =\frac{\mu^2}{{\sigma^{2}+\mu^2}}
\end{equation}

where, the function $F$ is defined by:

\begin{equation}
F(x) =\frac{{\Gamma(\frac{2}{x}) }^2}{\Gamma(\frac{1}{x})\Gamma(\frac{3}{x})}
\end{equation}

and $\alpha$ is defined by:

\begin{equation}
\alpha=\frac{ {(\sigma^{2}+\mu^2)} \Gamma(\frac{1}{\beta})}{\Gamma(\frac{3}{\beta})}
\end{equation}

$\mu$ is equal to the ordinary mean, since the histogram function of a picture is defined for positive values of the intensity levels. The entropy (in bits/pixel) is given by the formula:

\begin{equation}
H=2K \frac{\alpha}{\beta}\frac{(-\ln K)\Gamma(\frac{1}{\beta})+\Gamma(1+\frac{1}{\beta})}{\ln 2}
\end{equation}

In Fig.4 we display the variation of $H$ as a function of $\alpha$ in the interval [0.1-5] for a given $\beta$ picked in the interval [0.8-1.5]. Similarly, we show in Fig.5 the variation of $H$ as a function of $\beta$ in the interval [0.1-5] for a given  $\alpha$ picked in the interval [0.8-1.5].\\

Once the parameters $\alpha$ and  $\beta$ have been determined for a particular image along with its corresponding entropy, we proceed to the quantization procedure. We have implemented the simple Lloyd-Max algorithm [2,3] that minimizes the quantization error for the stretched exponential pdf. Figures 2 and 3 illustrate the reconstruction levels versus the decision levels in the case the number of bits is 6 and the values $\alpha= 1.2$ and $\beta= 1.56$. The quantized levels are compared to the Gaussian and Laplace cases for the same number of bits. In the stretched exponential case, an arbitrary value of $\alpha$ is valid in order to have  $\mu=0$ and $\sigma=1$. Choosing $\alpha=1.2$ as in the one-sided Gamma pdf, $\beta$ is obtained as a solution of Eq.17.

As a specific illustration of our methodology, we take two medical images (MRI-T1 format images of human brain sections) and apply the above analysis.
The histograms of the images and the Stretched exponential fits are shown in figures 6 and 7, whereas the values of $\alpha$ and  $\beta$ along with the entropies are given in Table 10. \\

\section{Conclusion}
We have developed an Lloyd-Max type algorithm  to build several quantization tables for
arbitrary pdf's. We have considered symmetric and single-sided pdf's. Our results
agree with published values in the 4-bit quantization case and outperform them in 
accuracy.  In the case of medical images, we develop a fitting procedure to the Stretched exponential case by taking raw histograms of the brain MRI images.

{\bf Acknowledgments}\\
This work started while the author was with the Department of  Electrical Engineering 
and with TRLabs in Saskatoon, Canada. The  author wishes to thank Kunio Takaya for many enlightening discussions. This work was supported in part by a Canada NSERC University fellowship grant.

\vspace{1cm}
\centerline{\Large\bf Table Captions}
\vspace{1cm}

{\bf Note:} Tables 1-7 is a set of tables displaying 4-bit quantization levels for various pdf's. Decision levels and Reconstruction levels according to the Lloyd-Max algorithm. The calculated 4-bit quantization decision and reconstruction levels use the following double-sided pdf's with mean $\mu=0$  and standard deviation $\Delta x=1$. The first set of Table 1 contains the following two-sided symmetric pdf's:
\begin{enumerate}
\item  Uniform pdf.
\item Gaussian pdf.
\item Laplace pdf.
\item Double-sided Gamma pdf.
\item Stretched Exponential pdf ($\alpha=1.2$) 

The second set contains one-sided pdf's with mean and standard deviation $\Delta x=1$ given below:

\item One-sided Rayleigh pdf:
Mean $\mu=\sigma \sqrt{\pi /2}$, Standard deviation: $\Delta x=\sigma \sqrt{2(1-\pi/4)}$.
\item One-sided Gamma pdf ($\alpha=1.2$):
Mean $\mu=\alpha \beta$, Standard deviation: $\beta \sqrt{\alpha}$.
\end{enumerate}

\begin{table}[!ht]
\begin{center}
\begin{tabular}{ ||c|c||}
\hline
$d_{i}$  &  $r_{i}$ \\
\hline
0.            &     0.108253  \\
0.216506      &     0.324760  \\
0.433013      &     0.541266   \\
0.649519      &     0.757772  \\
0.866025      &     0.974279  \\
1.08253      &     1.19079   \\
1.29904       &     1.40729 \\
1.51554      &      1.62380  \\
1.73205      &             \\
\hline
\end{tabular}
\caption{Uniform pdf (Double-sided Symmetrical, mean=0, Standard deviation=1). All values agree with corresponding entries of Table 9.}
\end{center}
\label{tab1}
\end{table}

\begin{table}[!ht]
\begin{center}
\begin{tabular}{ ||c|c||}
\hline
$d_{i}$  &  $r_{i}$ \\
\hline
0.         &    0.128481 \\
0.258396   &    0.388311 \\
0.522764   &    0.657218 \\
0.800124   &    0.943031 \\
1.10013    &    1.25723  \\
1.43836    &    1.61949  \\
1.84537    &    2.07126  \\
2.40409   &     2.73693  \\
$\infty$    &           \\
\hline
\end{tabular}
\caption{ Gaussian ( Double-sided Symmetrical, mean=0, Standard deviation=1).
All values agree with corresponding entries of Tables 8 and 9.}
\end{center}
\label{tab2}
\end{table}

\begin{table}[!ht]
\begin{center}
\begin{tabular}{ ||c|c||}
\hline
$d_{i}$  &  $r_{i}$ \\
\hline
        0.	     &  0.124026 \\
	0.264503    &  0.404980  \\
	0.566953   & 	0.728925  \\
	0.920184   & 	1.11144  \\
	1.34498	   & 	1.57852   \\
	1.87852	   & 	2.17852  \\
	2.59871	   & 	3.01891  \\
	3.72726	   & 	4.43562  \\
       $\infty$   &       \\

\hline
\end{tabular}
\caption{ Laplacian pdf (Double-sided Symmetrical, mean=0, Standard deviation=1).
All values agree with corresponding entries of Tables 8 and 9.}
\end{center}
\label{tab3}
\end{table}

\begin{table}[!ht]
\begin{center}
\begin{tabular}{ ||c|c||}
\hline
$d_{i}$  &  $r_{i}$ \\
\hline
0.		   &  0.128894   \\
	0.261969	  &   0.395045  \\
	0.536929   & 	0.678813   \\
	0.834667    &  0.990521  \\
	1.16796	   & 	1.34541  \\
	1.55782	   & 	1.77023  \\
	2.04637	   &   2.32251  \\
	2.75049	   & 	3.17848   \\
       $\infty$   &       \\

\hline
\end{tabular}
\caption{ Strectched exponential pdf (Symmetrical, mean=0, Standard deviation=1) [$\alpha$=1.2, $\beta$=1.55622]}
\end{center}
\label{tab4}
\end{table}

\begin{table}[!ht]
\begin{center}
\begin{tabular}{ ||c|c||}
\hline
$d_{i}$  &  $r_{i}$ \\
\hline
0.          &   $7.25920 \hspace{1mm} 10^{-2}$ \\
0.229727    &   0.386861  \\
0.590566   &    0.794270  \\
1.05014    &    1.30602  \\
1.63150    &    1.95698   \\
2.38795    &    2.81892  \\
3.43725    &    4.05558  \\
5.11878	   &    6.18198  \\
$\infty$   &       \\
\hline
\end{tabular}
\caption{ Symmetric Gamma pdf (Double-sided Symmetrical, mean=0, Standard deviation=1). All values agree with corresponding entries of Table 9. Note that some misprints and missing endpoint entries exist in Table 9.}
\end{center}
\label{tab5}
\end{table}

\begin{table}[htbp]
\begin{center}
\begin{tabular}{ ||c|c||}
\hline
$d_{i}$  &  $r_{i}$ \\
\hline
0.         &  0.305692  \\
0.460637   &  0.615582  \\
0.750951   &  0.886320  \\
1.01303    &  1.13973  \\
1.26238    &  1.38502  \\
1.50639    &  1.62775   \\
1.74994    &  1.87214  \\
1.99707    &  2.12200  \\
2.25168    &  2.38136  \\
2.51817   &   2.65498  \\
2.80205   &   2.94913  \\
3.11095    &  3.27278  \\
3.45648   &   3.64019  \\
3.85856   &   4.07693  \\
4.35750   &   4.63807  \\
5.06416   &   5.49025  \\
$\infty$   &      \\
\hline
\end{tabular}
\caption{ Rayleigh pdf (Single-sided, Standard deviation=1). All values agree with corresponding entries of Table 8.}
\end{center}
\label{tab6}
\end{table}

\begin{table}[htbp]
\begin{center}
\begin{tabular}{ ||c|c||}
\hline
$d_{i}$  &  $r_{i}$ \\
\hline
0.		   & 	0.113691  \\
	0.215704  & 	0.317717  \\
	0.422980   & 	0.528243  \\
	0.639068   & 	0.749892  \\
	0.867942   &  0.985993  \\
	1.11297	    &  1.23994   \\
	1.37782	  & 	1.51571  \\
	1.66701	   & 	1.81830  \\
	1.98632	   & 	2.15434  \\
	2.34364	   & 	2.53293 \\
	2.75015	   & 	2.96737 \\
	3.22270	   & 	3.47803  \\
	3.78839	   & 	4.09876  \\
	4.49562	   & 	4.89248  \\
	5.44523	   & 	5.99798  \\
	6.92198	   & 	7.84597  \\
  $\infty$      &        \\
\hline
\end{tabular}
\caption{ One-sided Gamma pdf (Single-sided, Standard deviation=1) [$\alpha=1.2, \beta=0.912871$]}
\end{center}
\label{tab7}
\end{table}

\newpage

\begin{widetext}

\begin{table}[!ht]
\begin{center}
\begin{tabular}{ |c || c | c || c | c || c  | c || c | c ||}
\hline
      &  \multicolumn{2}{|c|}{Uniform}   &   \multicolumn{2}{|c|}{Gaussian}  &  \multicolumn{2}{|c|}{Laplacian}   &  \multicolumn{2}{|c|}{Rayleigh} \\ 
\hline
Bits  &  $d_{i}$  &  $r_{i}$  &  $d_{i}$  &  $r_{i}$  &  $d_{i}$  &  $r_{i}$  &  $d_{i}$  &  $r_{i}$ \\
\hline
1  &   -1.0000   &     -0.5000  &  $-\infty$   &  -0.7979  &  $-\infty$   &   -0.7071   &  0.0000   &  1.2657 \\ 
  &   0.0000   &  0.5000   &    0.0000  &   0.7979   &  0.0000   &     0.7071   &  2.0985  &   2.9313  \\
  &   1.0000  &   & $\infty$  &   & $\infty$  &   & $\infty$ & \\
\hline
2  &   -1.0000  &  -0.7500  &  $-\infty$  &  -1.5104  &  $-\infty$  &  -1.8340  &  0.0000  &  0.8079  \\
  &   -0.5000  &  -0.2500  &  -0.9816  &  -0.4528  &  -1.1269  &  -0.4198  &  1.2545  &  1.7010 \\
  &   -0.0000  &  0.2500  &  0.0000  &  0.4528  &  0.0000  &  0.4198  &  2.1667  &  2.6325 \\
  &   0.5000  &  0.7500  &  0.9816  &  1.5104  &  1.1269  &  1.8340  &  3.2465  &  3.8604 \\
  &   1.0000  &   & $\infty$  &   & $\infty$  &   & $\infty$ & \\
\hline
3  &   -1.0000  &  -0.8750  &  $-\infty$  &  -2.1519  &  $-\infty$  &  -3.0867  &  0.0000  &  0.5016  \\
  &   -0.7500  &  -0.6250  &  -1.7479  &  -1.3439  &  -2.3796  &  -1.6725  &  0.7619  &  1.0222 \\
  &   -0.5000  &  -0.3750  &  -1.0500  &  -0.7560  &  -1.2527  &  -0.8330  &  1.2594  &  1.4966 \\
  &   -0.2500  &  -0.1250  &  -0.5005  &  -0.2451  &  -0:5332  &  -0.2334  &  1.7327  &  1.9688 \\
  &   0.0000  &  0.1250  &  0.0000  &  0.2451  &  0.0000  &  0.2334  &  2.2182  &  2.4675 \\
  &   0.2500  &  0.3750  &  0.5005  &  0.7560  &  0.5332  &  0.8330  &  2.7476  &  3.0277 \\
  &   0.5000  &  0.6250  &  1.0500  &  1.3439  &  1.2527  &  1.6725  &  3.3707  &  3.7137 \\
  &   0.7500  &  0.8750  &  1.7479  &  2.1519  &  2.3796  &  3.0867  &  4.2124  &  4.7111 \\
  &   1.0000  &   & $\infty$  &   & $\infty$  &   & $\infty$ & \\
\hline
4  &   -1.0000  &  -0.9375  &  $-\infty$  &  -2.7326  &  $-\infty$  &  -4.4311  &  0.0000  &  0.3057 \\ 
  &   -0.8750  &  -0.8125  &  -2.4008  &  -2.0690  &  -3.7240  &  -3.0169  &  0.4606  &  0.6156 \\
  &   -0.7500  &  -0.6875  &  -1.8435  &  -1.6180  &  -2.5971  &  -2.1773  &  0.7509  &  0.8863 \\
  &   -0.6250  &  -0.5625  &  -1.4371  &  -1.2562  &  -1.8776  &  -1.5778  &  1.0130  &  1.1397 \\
  &   -0.5000  &  -0.4375  &  -1.0993  &  -0.9423  &  -1.3444  &  -1.1110  &  1.2624  &  1.3850 \\
  &   -0.3750  &  -0.3125  &  -0.7995  &  -0.6568  &  -0.9198  &  -0.7287  &  1.5064  &  1.6277 \\
  &   -0.2500  &  -0.1875  &  -0.5224  &  -0.3880  &  -0.5667  &  -0.4048  &  1.7499  &  1.8721 \\
  &   -0.1250  &  -0.0625  &  -0.2582  &  -0.1284  &  -0.2664  &  -0.1240  &  1.9970  &  2.1220 \\
  &   0.0000  &  0.0625  &  0.0000  &  0.1284  &  0.0000  &  0.1240  &  2.2517  &  2.3814 \\
  &   0.1250  &  0.1875  &  0.2582  &  0.3880  &  0.2644  &  0.4048  &  2.5182  &  2.6550 \\
  &   0.2500  &  0.3125  &  0.5224  &  0.6568  &  0.5667  &  0.7287  &  2.8021  &  2.9492 \\
  &   0.3750  &  0.4375  &  0.7995  &  0.9423  &  0.9198  &  1.1110  &  3.1110  &  3.2729 \\
  &   0.5000  &  0.5625  &  1.0993  &  1.2562  &  1.3444  &  1.5778  &  3.4566  &  3.6403 \\
  &   0.6250  &  0.6875  &  1.4371  &  1.6180  &  1.8776  &  2.1773  &  3.8588  &  4.0772 \\
  &   0.7500  &  0.8125  &  1.8435  &  2.0690  &  2.5971  &  3.0169  &  4.3579  &  4.6385 \\
  &   0.8750  &  0.9375  &  2.4008  &  2.7326  &  3.7240  &  4.4311  &  5.0649  &  5.4913 \\
  &   1.0000  &   & $\infty$  &   & $\infty$  &   & $\infty$ & \\
\hline

\end{tabular}
\end{center}
\label{tab8}
\caption{Published reconstruction and decision levels for Uniform, Gaussian, Laplace and Rayleigh pdf's. The number of quantization bits ranges from 1 to 4. All double-sided pdf's are with mean $\mu=0$ and standard deviation $\Delta x=1$. The table also contains the decision and reconstruction levels for the one-sided Rayleigh pdf with standard deviation $\Delta x=1$. Adapted from \cite{pratt}.}
\end{table}

\begin{table}[!ht]
\begin{center}
\begin{tabular}{ |c | c || c | c || c | c || c  | c || c | c ||}
\hline
Bits   &    &   \multicolumn{2}{|c|}{1}   & \multicolumn{2}{|c|}{2}  &  \multicolumn{2}{|c|}{3}  &  \multicolumn{2}{|c|}{4} \\
\hline
pdf  &  j  &  $d_{j}$  &  $r_{j}$   &  $d_{j}$  &  $r_{j}$  &  $d_{j}$  &  $r_{j}$  &  $d_{j}$  &  $r_{j}$\\
\hline
     & 1   &  0.000   & 0.866 & 0.000  & 0.433 & 0.000 & 0.217 & 0.000 & 0.109  \\
     & 2  &   &   & 0.866  & 1.299 & 0.433 & 0.650  &  0.217 & 0.326\\
     & 3  &  &  &  &  & 0.866 & 1.083 & 0.433 & 0.542\\
  U  & 4 &  &  &  &  & 1.299 & 1.516 & 0.650 & 0.759\\
     & 5 &  &  &  &  &  &  & 0.866 & 0.975\\
     & 6 &  &  &  &  & & & 1.083 & 1.192\\
     & 7 &  &  &  &  & & & 1.299 & 1.408\\
     & 8 &  &  &  &  & & & 1.516 & 1.624\\
\hline
  & 1 & 0.000  & 0.798 & 0.000  & 0.453 & 0.000 & 0.245 & 0.000 & 0.128  \\
  & 2 &   &   & 0.982  & 1.510 & 0.501 & 0.756 & 0.258 & 0.388  \\
  & 3  &  &  &  &  & 1.050 & 1.344 & 0.522 & 0.657\\
  & 4 &  &  &  &  &  1.748 & 2.152 & 0.800 & 0.942 \\
G  & 5 &  &  &  &  &  &  & 1.099 & 1.256  \\
  & 6 &  &  &  &  & & & 1.437 & 1.618  \\
  & 7 &  &  &  &  & & & 1.844 & 2.069  \\
  & 8 &  &  &  &  & & & 2.401 & 2.733  \\
\hline
  & 1 & 0.000  & 0.707 & 0.000  & 0.420 & 0.000 & 0.233 & 0.000 & 0.124  \\
  & 2 &   &   & 1.127  & 1.834 & 0.533 & 0.833 & 0.264 & 0.405  \\
  & 3 &  &  &  &  & 1.253 & 1.673 & 0.567 & 0.729  \\
  & 4 &  &  &  &  & 2.380 & 3.087 & 0.920 & 1.111  \\
 L  & 5 &  &  &  &  &  &  & 1.345 & 1.578  \\
  & 6 &  &  &  &  &  &  & 1.878 & 2.178  \\
  & 7 &  &  &  &  &  &  & 2.597 & 3.017  \\
  & 8 &  &  &  &  &  &  & 3.725 & 4.432  \\
\hline
   & 1 & 0.000  & 0.577 & 0.000  & 0.313 & 0.000 & 0.155 & 0.000 & 0.073  \\
   & 2 &  &  & 1.268  & 2.223 & 0.527 & 0.899 & 0.230 & 0.387  \\
   & 3 &  &  &  &  & 1.478 & 2.057 & 0.591 & 0.795  \\
 $\Gamma$  & 4 &  &  &  &  & 3.089 & 4.121 & 0.051 & 1.307 \\
   & 5 &  &  &  &  &  &  & 1.633 & 1.959 \\
   & 6 &  &  &  &  &  &  & 1.390 & 2.822 \\
   & 7 &  &  &  &  &  &  & 3.422 & 4.061 \\
   & 8 &  &  &  &  &  &  & 5.128 & 6.195 \\
\hline
\end{tabular}
\end{center}
\label{tab9}
\caption{Published values of reconstruction and decision levels for Symmetric Double-sided Uniform, Gaussian, Laplace and Gamma pdf's. The number of quantization bits ranges from 1 to 4. Only positive values are given. Several misprints and missing entries are found in the table. For instance in the Uniform case the last entry 1.73205 (given in Table 1) is missing. The entry $\infty$ is missing in the Laplace case. In the Double-sided Gamma pdf, the entries 0.051 and 1.390 are wrong as shown in Table 5. The last entry $\infty$ is also missing. Adapted from \cite{jayant}.}
\end{table}

\newpage 

\begin{table}[!ht]
\begin{center}
\begin{tabular}{ |c || c | c|}
\hline
   & First Image  &  Second Image \\  \hline
Raw  Characteristics &   &  \\ 
\hline
Intensity Levels  & 	256    &   256 \\
Ordinary Mean   & 		5.19436  &  5.47491 \\
Standard Deviation   & 		8.98966    &  8.53209 \\
Variance  &	80.8140         &    72.7966  \\
Entropy/Pixel &		5.86363 bits &   6.09553 bits. \\
\hline\hline
Fitted histogram &  &  \\
\hline
$\alpha$ 	&	0.164285  &  0.655837  \\
$\beta$   & 	0.425672  & 0.486307 \\
Entropy/Pixel &   3.28169 bits  & 4.43392 bits\\
\hline
\end{tabular}
\end{center}
\label{tab10}
\caption{ Characteristics of the MRI images and values of $\alpha$ and $\beta$ from the stretched exponential fit to the histograms along with the entropies from the raw histograms.}
\end{table}

\end{widetext}

\vspace{1cm}
\centerline{\Large\bf Figure Captions}
\vspace{1cm}

\begin{itemize}

\item[Fig.\ 1:] Behaviour of the quantization error versus the first reconstructed level $r_{0}$. The presence of several local minima surrounded by large values of the error is the reason why the search for the global minimum is difficult when the starting guess value is far away.

\item[Fig.\ 2:] Comparative decision and reconstruction levels for 6-bit quantization. Two-sided symmetric pdf case (Uniform, Gamma, Stretched Exponential, Laplace and Gaussian).  All pdf's have zero mean and standard deviation $\Delta x=1$.

\item[Fig.\ 3:] Comparative decision and reconstruction levels for 6-bit quantization. One-sided pdf case (Rayleigh and Gamma). All pdf's have a standard deviation $\Delta x=1$.

\item[Fig.\ 4:] Entropy $H$ of the stretched exponential pdf as a function of $\alpha$ in the interval [0.1-5] for a given $\beta$ picked in the interval [0.8-1.5].

\item[Fig.\ 5:] Entropy $H$ of the stretched exponential pdf  as a function of $\beta$ in the interval [0.1-5] for a given $\alpha$ picked in the interval [0.8-1.5].

\item[Fig.\ 6:] First MRI image histogram and stretched exponential fit.

\item[Fig.\ 7:] Second MRI image histogram and stretched exponential fit.

\item[Fig.\ 8:] First analysed MRI medical image.

\item[Fig.\ 9:] Second analysed MRI medical image.

\end{itemize}

\begin{widetext}

\begin{figure}[htbp]
\begin{center}
\scalebox{0.7}{\includegraphics[angle=-90]{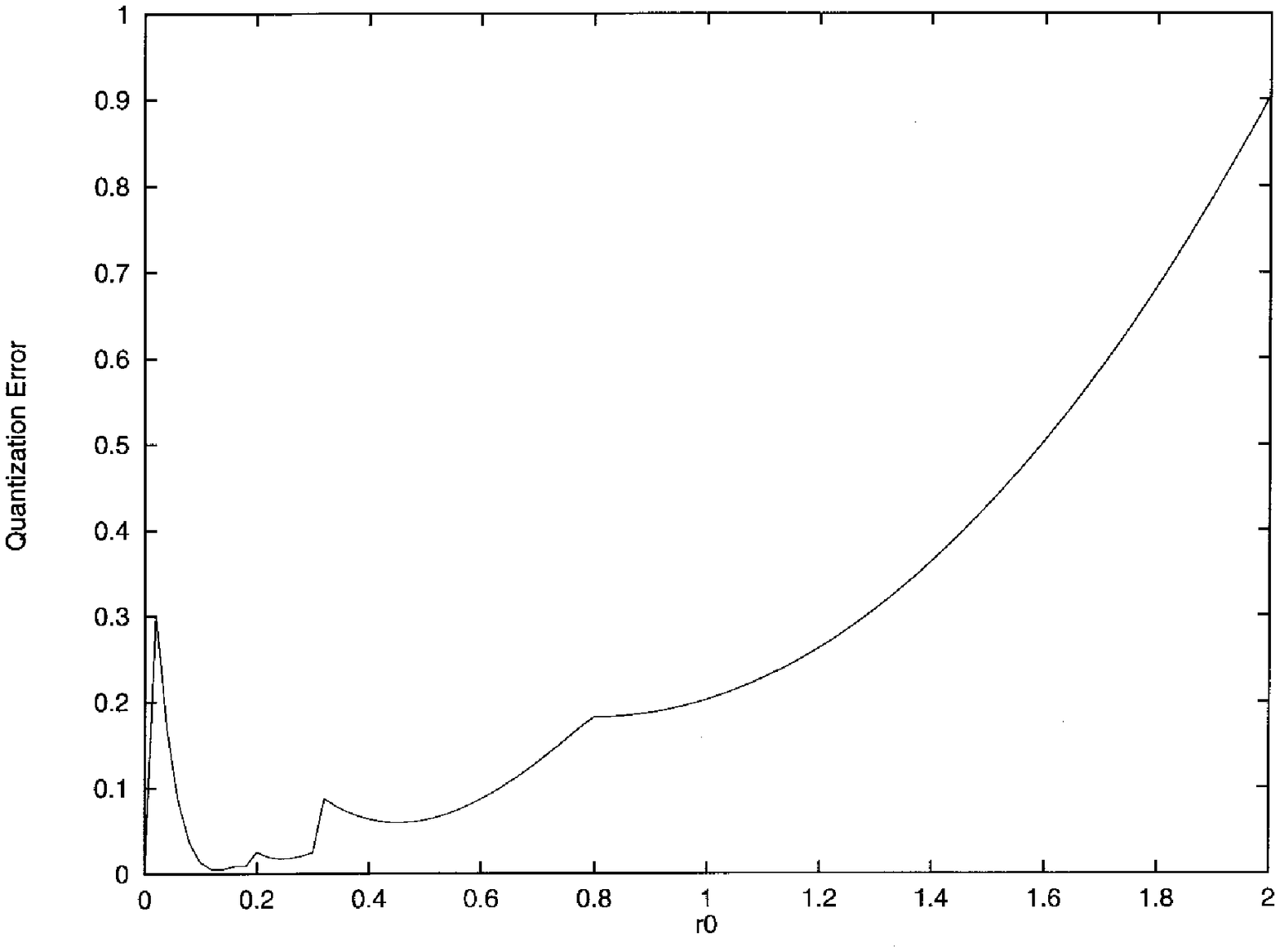}}
\caption{Behaviour of the quantization error versus the first reconstructed level $r_{0}$. The presence of several local minima surrounded by large values of the error is the reason why the search for the global minimum is difficult when the starting guess value is far away.} \label{fig1}
\end{center}
\end{figure}

\begin{figure}[htbp]
\begin{center}
\scalebox{0.6}{\includegraphics[angle=-90]{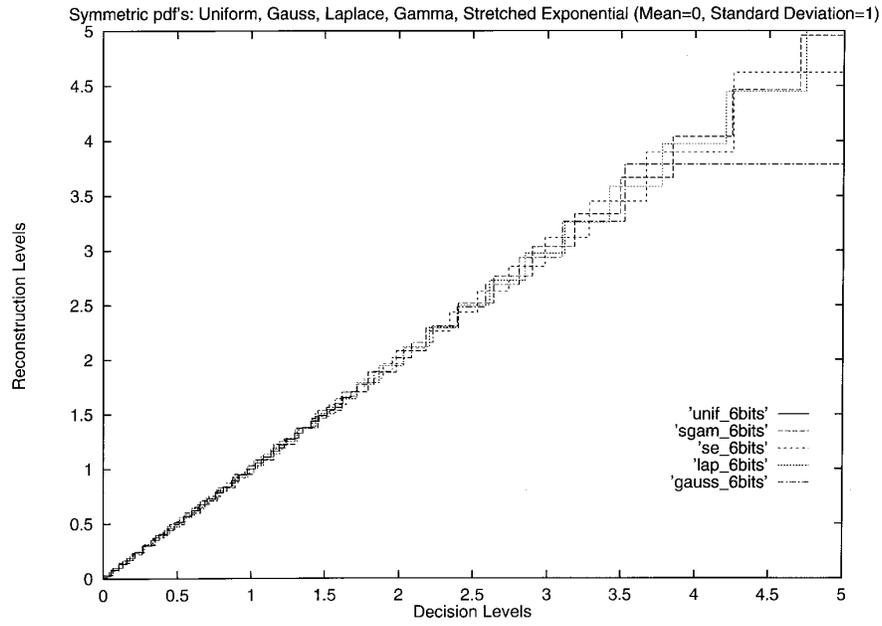}}
\caption{Comparative decision and reconstruction levels for 6-bit quantization. Two-sided symmetric pdf case (Uniform, Gamma, Stretched Exponential, Laplace and Gaussian).  All pdf's have zero mean and standard deviation $\Delta x=1$.}\label{fig2}
\end{center}
\end{figure}

\begin{figure}[htbp]
\begin{center}
\scalebox{0.6}{\includegraphics[angle=-90]{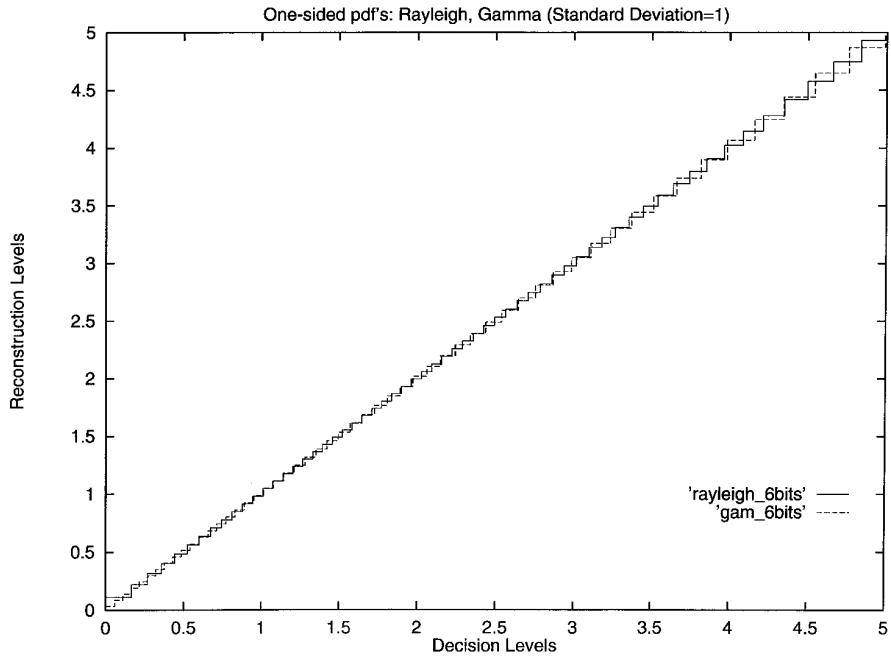}}
\caption{Comparative decision and reconstruction levels for 6-bit quantization. One-sided pdf case (Rayleigh and Gamma). All pdf's have a standard deviation $\Delta x=1$.} \label{fig3}
\end{center}
\end{figure}

\begin{figure}[htbp]
\begin{center}
\scalebox{0.6}{\includegraphics[angle=-90]{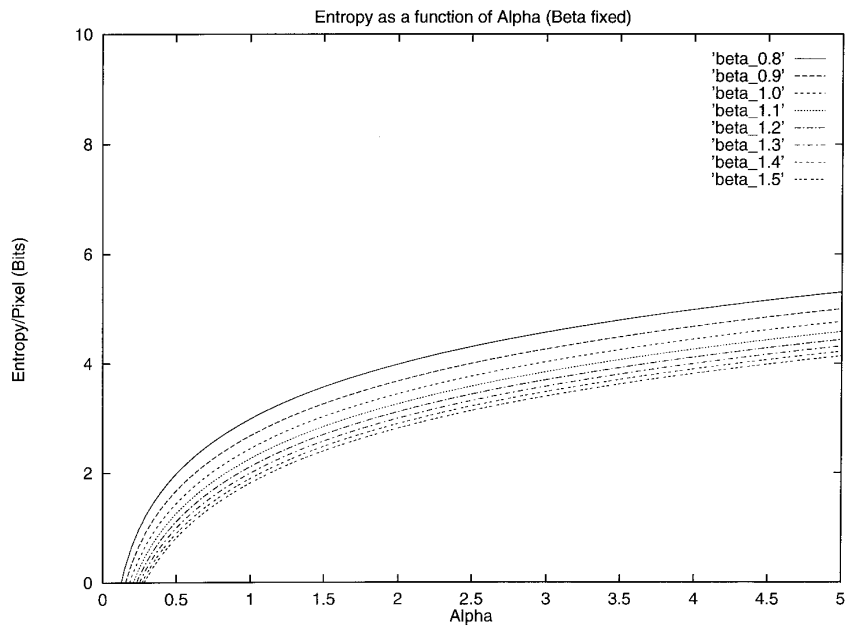}}
\caption{ Entropy $H$ of the stretched exponential pdf as a function of $\alpha$ in the interval [0.1-5] for a given $\beta$ picked in the interval [0.8-1.5]. The uppermost curve corresponds to $\beta=0.8$ and the lowest is for $\beta=1.5$.} \label{fig4}
\end{center}
\end{figure}

\begin{figure}[htbp]
\begin{center}
\scalebox{0.6}{\includegraphics[angle=-90]{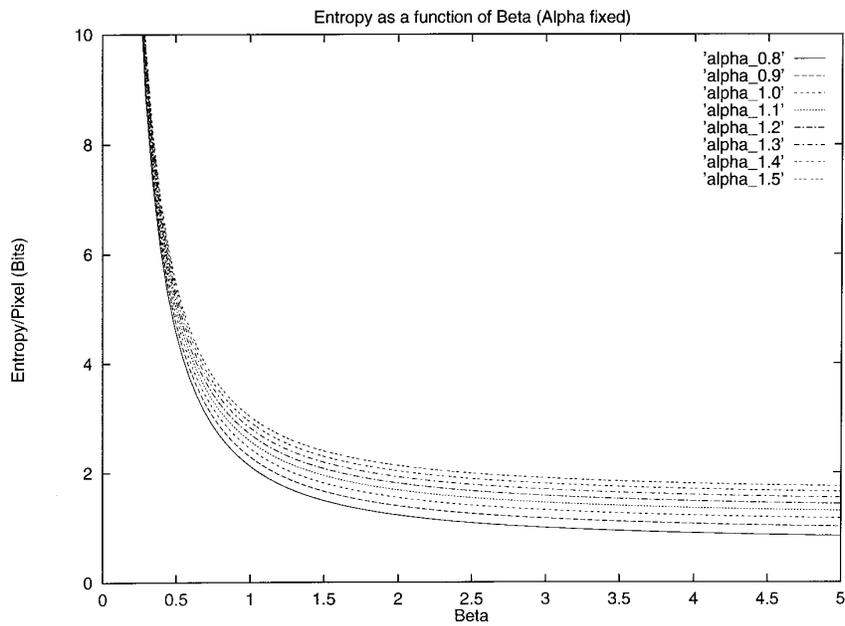}}
\end{center}
\caption{Entropy $H$ of the stretched exponential pdf  as a function of $\beta$ in the interval [0.1-5] for a given $\alpha$ picked in the interval [0.8-1.5]. The top curve corresponds to $\alpha=0.8$ and the lowest is for $\alpha=1.5$.} \label{fig5}
\end{figure}

\begin{figure}[htbp]
\begin{center}
\scalebox{0.6}{\includegraphics[angle=-90]{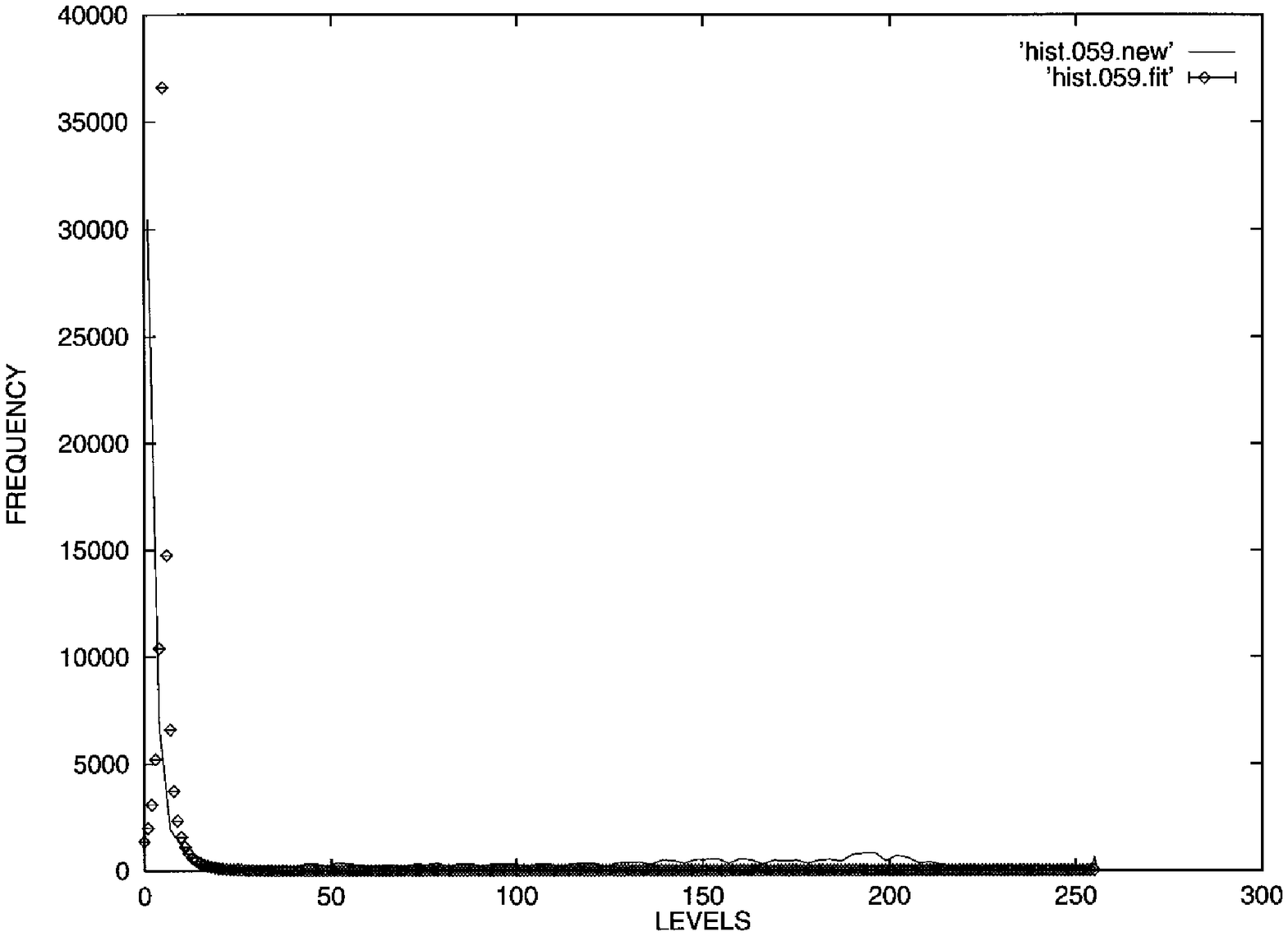}}
\caption{First MRI image histogram and stretched exponential fit. The fit parameters are displayed in Table X. The raw histogram is indicated with a continuous line and the diamonds $\Diamond$ correspond to the fit.} \label{fig6}
\end{center}
\end{figure}

\begin{figure}[htbp]
\begin{center}
\scalebox{0.7}{\includegraphics[angle=-90]{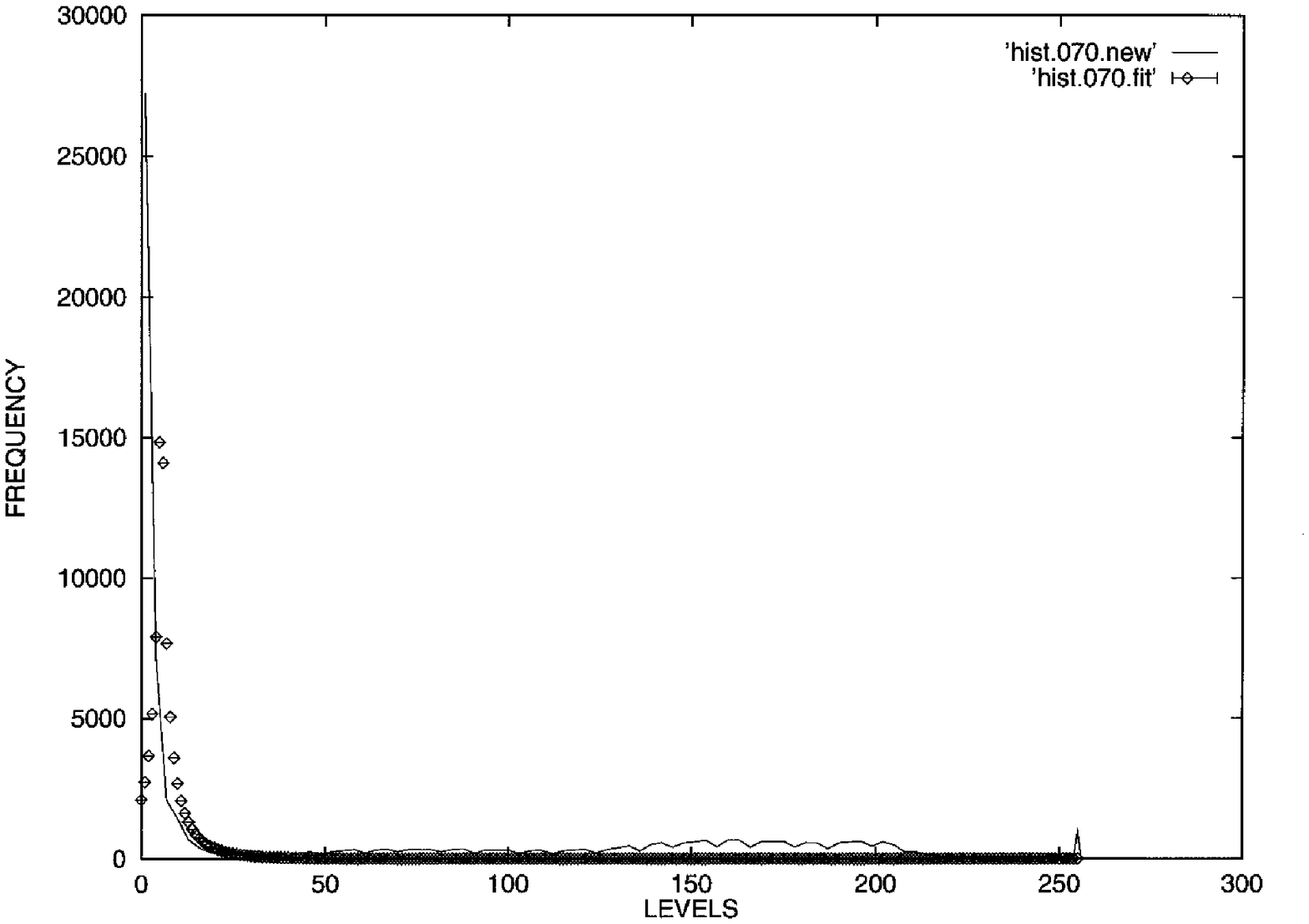}}
\caption{Second MRI image histogram and stretched exponential fit. The fit parameters are displayed in Table X.  The raw histogram is indicated with a continuous line and the diamonds $\Diamond$ correspond to the fit.} \label{fig7}
\end{center}
\end{figure}

\begin{figure}[htbp]
\begin{center}
\includegraphics{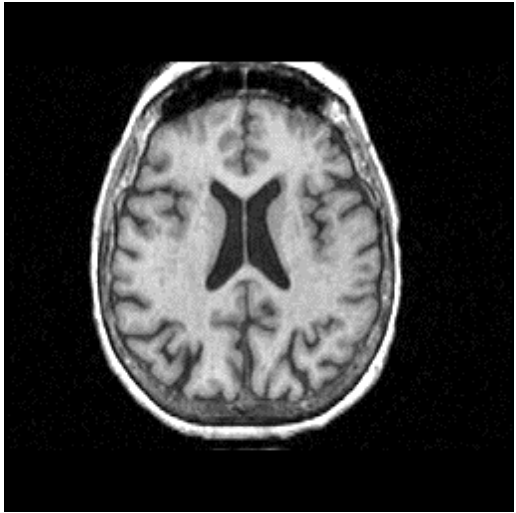}
\caption{First analysed MRI medical image.} \label{fig8}
\end{center}
\end{figure}

\begin{figure}[htbp]
\begin{center}
\includegraphics{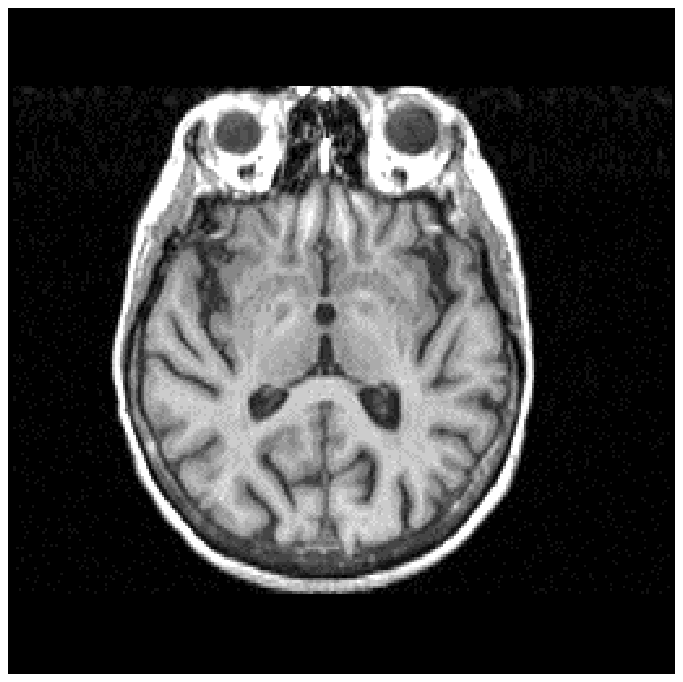}
\caption{Second analysed MRI medical image.} \label{fig9}
\end{center}

\end{figure}

\end{widetext}

\end{document}